\documentclass[10pt,twocolumn]{IEEEtran}
\usepackage{amsmath,amssymb,mathrsfs}
\usepackage{color,times}
\usepackage{epsfig}
\usepackage{amsmath}
\usepackage{amssymb}
\usepackage{algorithm,algpseudocode}

\begin{document}

\title{Study of Polarization-Driven Shortening for Polar Codes Designed with the Gaussian Approximation}

\author{Robert M. Oliveira and Rodrigo C. de Lamare
\thanks{Robert M. Oliveira and Rodrigo C. de Lamare - Centre for Telecommunications
Studies (CETUC), Pontifical Catholic University of Rio de Janeiro
(PUC-Rio), Rio de Janeiro-RJ, Brasil, E-mails: rbtmota@gmail.com e
delamare@cetuc.puc-rio.br} }

\maketitle


\begin{abstract}
This paper presents a polarization-driven (PD) shortening technique
for the design of rate-compatible polar codes. The proposed
shortening strategy consists of reducing the generator matrix by
relating its row index with the channel polarization index. We
assume that the shortened bits are known by both the encoder and the
decoder and employ successive cancellation (SC) for decoding the
shortened codes constructed by the proposed PD technique. A
performance analysis is then carried out based on the Spectrum
Distance (SD). Simulations show that the proposed PD-based shortened
polar codes outperform existing shortened polar codes.
\end{abstract}


\section{Introdution}

Polar codes, proposed by Arikan, are low-complexity
capacity-achieving codes based on the phenomenon called channel
polarization \cite{Arikan}. A typical construction of conventional polar codes
is based on the Kronecker product, which is restricted to the
lengths $2^l$ $(l = 1,2,...)$. Polar codes with arbitrary lengths
can be obtained by shortening or puncturing \cite{Bioglio}, which will be
required for 5G scenarios, where code lengths ranging from $420$ to
$1920$ bits with various rates will be adopted \cite{Osseiran} and \cite{Ferreira}.
Shortened and punctured polar codes  can be decoded in a similar way
to conventional polar codes.

Various shortening and puncturing methods for polar codes have been
proposed in the literature \cite{Niu}-\cite{Wang} and evaluated with
successive cancellation (SC) or belief propagation (BP) decoding. In
\cite{Hong}-\cite{NiuChen} puncturing methods have been reported
using BP decoding based on optimization techniques employing
retransmission schemes such as Hybrid Automatic Repeat reQuest
(HARQ). Different properties of punctured codes have been explored:
minimum distance, exponent binding, stop tree drilling, and the
reduced generating matrix method \cite{Chen}-\cite{Eslami}. Schemes
that depend on the analysis of density evolution were proposed in
\cite{EslamiPishro} and \cite{Kim}. On the other hand, shortening
methods have been studied with SC decoding. With shortening
techniques, we freeze a bit channel that receives a fixed zero
value. The decoder, however, uses a plus infinity log-likelihood
ratio (LLR) for that code bit as it is often assumed that this value
is known. The study in \cite{Miloslavskaya} proposed a search
algorithm to jointly optimize the shortening patterns and the values
of the shortened bits. The work in \cite{Wang} devised a simple
shortening method, reducing the generator matrix based on the weight
of the columns (CW). In \cite{Niu} the reversal quasi-uniform
puncturing scheme (RQUP) for reducing the generator matrix has been
proposed.

In this paper, we propose a polarization-driven (PD) shortening
technique based on the Gaussian Approximation (GA) \cite{Trifonov},
where the channel polarization index determines the channel
shortening patterns \cite{pd_short}. In particular, we describe the
design of rate-compatible polar codes using the PD method and its
application to fifth generation (5G) wireless system scenarios. We
also carry out an analysis of the proposed PD method using the
Spectrum Distance (SD) \cite{Niu} and assess the performance of
design examples via simulations.

\section{System Model and Problem Statement}

Fig.1 shows a block diagram of the polar coding system considered
in this paper.

\begin{figure}[htb]
\begin{center}
\includegraphics[scale=0.6]{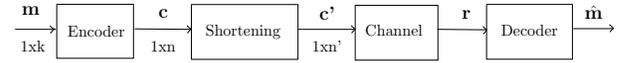} 
\caption{System Model.}
\end{center}
\label{figura:fig23b}
\end{figure}

In this system, $\textbf{m}$ is the binary message with $k$ bits
that is transmitted. The $n \times n$ generator matrix $\textbf{G}$
encodes the message $\textbf{m}$ and produces the codeword
$\textbf{c}$ with $n$ bits. With an appropriate shortening
technique, the codeword $\textbf{c}$ has its length reduced to $n'$,
resulting in the shortened codeword $\textbf{c'}$, where $2^{l-1} <
n' < 2^l$, where $l$ is an integer that defines the levels in the
polarization tree, $l=\log_2n$. The shortened codeword $\textbf{c'}$
is then transmitted over a channel with additive white Gaussian
noise (AWGN), resulting in the received vector
$\textbf{r}=\textbf{c'}+\textbf{w}$, where $\textbf{w}$ is the
vector corresponding to the noise. In the decoding step, the
decoding algorithm observes $\textbf{r}$ in order to estimate
$\textbf{m}$. We call it an estimated message $\hat{\textbf{m}}$,
and if $\textbf{m}=\hat{\textbf{m}}$ we say that the message has
been fully recovered. The problem we are interested in solving is
how to design shortened codes with the best performance.

\section{Polar Coding System}

Let $W:X \to Y$ denote a binary discrete memoryless channel (B-DMC),
with input alphabet $X=\{0,1\}$, output alphabet $Y$, and the
channel transition probability $W(y|x)$, $x \in X$, $y \in Y$. The
mutual information of the channel with equiprobable inputs, or
symmetric capacity, is defined by \cite{Arikan}
\begin{equation}
I(W) = \sum\limits_{y \in Y}\sum\limits_{x \in X}\frac{1}{2}W(y|x)log\frac{W(y|x)}{\frac{1}{2}W(y|0)+\frac{1}{2}W(y|1)}
\end{equation}
and the corresponding reliability metric, the Bhattacharyya
parameter, is described by [1]
\begin{equation}
Z(W) = Z_0 = \sum\limits_{y \in Y}\sqrt{W(y|0)W(y|1)}
\end{equation}
Applying the channel polarization transform for $n$ independent uses
of $W$ we obtain after channel combining and splitting operations
the group of polarized channels $W_n^{(i)}:X \to Y \times X^{i-1}$,
$i=1,2, \ldots ,n$. The channel polarization index $Z(W_n)$ over
AWGN channels is calculated using the GA method \cite{Trifonov} with
the following recursions:
\begin{equation}
\begin{cases}
Z(W^{(2i-1)}_n) = \phi^{-1}(1-(1-\phi(Z(W^{(i)}_{n/2})))^2)\\
Z(W^{(2i)}_n) = 2Z(W^{(i)}_{n/2}), \label{index}
\end{cases}
\end{equation}
where
\begin{equation}
\phi(x) \triangleq
\begin{cases}
\exp(-0.4527x^{(0.86)}+0.0218) \ {\rm if} \ 0 < x \leq 10\\
\sqrt{\frac{\pi}{x}}(1-\frac{10}{7x})\exp(\frac{-x}{4}) \ ~~~~~~~~~~~~~~ {\rm if} \ x > 10\\
\end{cases}
\end{equation}
In general, we use in the notation ${\bf a}^n_1$ to designate a
vector $(a_1,a_2,...,a_n)$ and $|{\bf a}^n_1|$ to refer to its
cardinality. The channel polarization theorem \cite{Arikan} states
that $I(W_n^{(i)})$ converges to either 0 (completely noisy
channels) or 1 (noiseless channels) as $n \to \infty$ and the
fraction of noiseless channel tends to $I(W)$, while polarized
channels converge to either $Z(W_n^{(i)})=1$ or $Z(W_n^{(i)})=0$.
The vector $\textbf{c}=(\textbf{u}_A, \textbf{u}_{A^c})$, where
$\textbf{m} = \textbf{u}_A$, for some $A \subset\{1,...,n\}$ denotes
the set of information bits and $A^c \subset \{1,...,n\}$ denotes
the set of frozen bits. We select the $|\textbf{m}|$ channels to
transmit information bits such that $Z(W_n^{(i)}) \leq
Z(W_n^{(j)})$. For encoding, a codeword is generated by
$\textbf{c}=c^n_1=\textbf{m}\textbf{B}_n\textbf{G}^{\otimes l}_2$,
where $\textbf{m}$ is the information sequence, $\textbf{B}_n$ is
the bit-reversal permutation matrix, $\otimes l$ is the $l$-th
Kronecker power and $\textbf{G}_2 =
\footnotesize\left[\begin{array}{cc}
1 & 0 \\
1 & 1 \end{array} \right]$ is the kernel matrix. We adopt the SC
decoder to estimate the information bits as \cite{Arikan}
\begin{equation}
\hat{u}_i=\arg\max\limits_{u_i \in \{0,1\}} \ W_n^{(i)}(y_1^n,u_1^{i-1}|u_i),i \in A
\end{equation}

\section{Proposed Polarization-Driven Shortening}

In this section, we detail the proposed PD shortening technique and
show how to calculate the polarization channels. Polar codes are
nonuniversal \cite{Arikan}, i.e., different polar codes are
generated depending on the specified value of the signal-to-noise
ratio (SNR), known as design-SNR. The design-SNR choice is critical
for ensuring good performance in all SNRs of interest and in this
work we adopt the design-SNR equal to zero.

The purpose of shortening is to reduce the size of the generator
matrix $\textbf{G}_n$ from $n \times n$ to $n' \times n'$, such that
$n' < n$. In particular, the size reduction is obtained by
eliminating rows and columns of the matrix $\textbf{G}_n$. Consider
the shortened vector $\textbf{p}$ which contains the indexes of the
rows of the matrix $\textbf{G}_n$ to be shortened, where
$|\textbf{p}|=n-n'$ indicates its number of elements. Consider the
set $\textbf{G}_n^r$ of all shortened matrices,
$r=(1,\ldots,\binom{n}{n'})$. We define
\begin{equation}
\textbf{G}_n^*=\arg \min\limits_{\textbf{G}^r} \ \rm BER,
\end{equation}
where the bit error rate (BER) is adopted. Many works resort to
exhaustive searches with an optimization method to determine
$\textbf{G}_n^*$, its the optimal shortened generator matrix. In
contrast to prior work, we propose the PD shortening technique for
computing $\textbf{p}$, where the channels with the lowest
polarization indexes are eliminated.

Using the notation in \cite{Arikan}, for $n = 8$ we have $l$ stages of polarization (\ref{index}) are
\begin{itemize}
 \item stage 1: \\ $Z(W^+)$ and $Z(W^-)$
 \item stage 2: \\ $Z(W^{++})$, $Z(W^{-+})$, $Z(W^{+-})$ and $Z(W^{--})$
 \item stage 3: \\ $Z(W^{+++}),\ Z(W^{-++}),\ Z(W^{+-+}),\ Z(W^{--+}),\\ Z(W^{++-}),\ Z(W^{-+-}),\ Z(W^{+--})\ \rm and\ Z(W^{---})$.
\end{itemize}

The channels $(W^{+++},W^{-++},W^{+-+},W^{--+},W^{++-},
\\W^{-+-},W^{+--},W^{---})$ can be written with $(W_0, W_1, W_2,\\
W_3, W_4, W_5, W_6, W_7)$. We define the polarization vector as
\begin{equation}
\textbf{b} \triangleq \left[\begin{array}{cccc} Z(W_0); Z(W_1); \ldots; Z(W_{n-1})
\end{array}\right]^T.
\end{equation}
As an example for stage 3 with normalized values, we have
$\textbf{b} =[0.992,0.882,0.915,0.578,0.938,0.639,0.715,0.000]^T$.
The key idea of the proposed PD method is to remove in the generator
matrix $\textbf{G}_n$ the rows that correspond to the channels with
smallest values of polarization.

These channels can be obtained by sorting the polarization vector
$\textbf{b}$. The goal of sorting is to determine a permutation
$k(1)k(2) \ldots k(n)$ of the indexes $\{1,2,...,n\}$ that will
organize the entries of the polarization vector $\textbf{b}$ in
increasing order \cite{Knuth}:
\begin{equation}
Z(W_{k(1)}) \leq Z(W_{k(2)}) \leq \ldots \leq Z(W_{k(n)}) \label{order}
\end{equation}
Consider the sort function $[\textbf{a}, \textbf{k}]= \rm
sort(\textbf{b})$ which implements (\ref{order}), where $\textbf{a}$
lists the sorted $\textbf{b}$ and $\textbf{k}$ contains the
corresponding indexes of $\textbf{a}$. Table 1 shows an example of
the polarization vector $\textbf{b}$ for $n = 8$, sorting vector
$\textbf{a}$ and the new index $\textbf{k}$.

\begin{table}[htb]
\vspace{-0.75 em} \caption{\label{tabela}Polarization vector $\rm
\textbf{b}$ for $n = 8$}
\begin{center}
\vspace{-0.75 em} \scriptsize
\begin{tabular}{|c||c|c|c|c|c|c|c|c|}\hline
\textit{$\rm
\textbf{b}$}&0.992&0.882&0.915&0.578&0.938&0.639&0.715&0.000\\\hline
\textit{$\rm index$}&1&2&3&4&5&6&7&8\\\hline
\multicolumn{9}{|c|}{After sorting} \\ \hline \textit{$\rm
\textbf{a}$}&0.000&0.578&0.639&0.715&0.882&0.915&0.938&0.992\\\hline
\textit{$\rm \textbf{k}$}&8&4&6&7&2&3&5&1\\\hline
\end{tabular}
\scriptsize \vspace{-0.5 em}
\end{center}
\end{table}

The vector $\textbf{k} = [8,4,6,7,2,3,5,1]$ contains the indexes of
the polarization values of the channels in increasing order, which
are used to obtain the shortening vector $\textbf{p}$ of the
proposed PD method:
\begin{equation}
\textbf{p}=[k(1), \ldots , k(n-n')],
\end{equation}
with $n-n'$ being the shortening length. In Algorithm 1 we have
included a pseudo-code of the proposed PD method with details of the
size reduction of the generator matrix $\textbf{G}_n$ and the
shortening of the channels with the lowest polarization values.

\begin{algorithm}
\caption{Proposed PD algorithm}
\begin{algorithmic}[1]
\State Given a shortened codeword with length $n'$

\State Use $\textbf{G}_n$ as the base matrix

\State Index each column by $\{1,2,...,n\}$

\State Index each row by $\{1,2,...,n\}$

\State Calculate the polarization channel vector $\textbf{b}$ for n

\State Calculate $[\textbf{a},\textbf{k}]= \rm sort(\textbf{b})$

\State Calculate the shortening vector $\textbf{p}=[k(1), \ldots ,
k(n-n')]$

\For{y = 1 to $|\textbf{p}|$}

\State $r_{\rm min} \leftarrow \textbf{p}(y)$

\State Delete row from $\textbf{G}_n$ with index $r_{\rm min}$
\State Delete column from $\textbf{G}_n$ with index $r_{\rm min}$
\EndFor

\State $\textbf{end for}$
\end{algorithmic}
\end{algorithm}

We consider now an example with shortened polar codes with length
$n'= 5$. For the shortening of $\textbf{G}_8$ to $\textbf{G}_5$, the
channels with the lowest polarization rank values are $W_8$, $W_4$
and $W_6$, the shortening vector is $\textbf{p}=(8,4,6)$ and
$|\textbf{p}|=3$. The $1$st element of $\textbf{p}$ is 8, which
results in the deletion of the $8$th column and the $8$th row of
$\textbf{G}_8$. The $2$nd element of $\textbf{p}$ is $4$, which
requires the elimination of the 4th column and the $4$th row. At
last, the $3$rd element of the $\textbf{p}$ is 6, which requires the
deletion of the $6$th column and the $6$th row. The matrix
$\textbf{G}_8$ with the indication of the deletions and the
resulting generator matrix $\textbf{G}_5$ are given by
As the code has been shortened, the reliability of the bit channels
changes and the information set should change accordingly. The
$Z(W)$ parameters of the polarized channels shortened are smaller
than those of the original polarized channels, we consider in this
paper that the order of channel polarization given by (8) does not
change after shortening. The shortened codeword $\textbf{c'}$
generated with $\textbf{G}_{n'}$, which contains the bits of the
binary message $\textbf{m}=\textbf{u}_A$ such that $Z(W_{n'}^{(i)})
\leq Z(W_{n'}^{(j)})$ for all $i \in A$, $j \in A^c$ and
$\textbf{u}_{A^c}=(u_i:i \in A^c|u_i=0)$ is then transmitted over a
channel. The PD shortening method assumes that the channels
remaining after shortening keep their polarization ordering.
Therefore, any rate for the shortening of the polar code can be
arbitrarily chosen as in conventional polar codes, where the
channels with the smallest polarization indexes are chosen for the
information bits.

\section{Analysis}

The work in \cite{Liu} examined maximum likelihood (ML) decoding for
polar codes and demonstrated that systematic coding yields better
BER performance than non-systematic coding with the same FER
performance for both encoding schemes. However, ML decoding is quite
costly since it compares all possible codewords for a given polar
code using the Hamming distance.

We employ the SD that has been studied in \cite{Niu} for analysis
due to its lower computational cost than ML decoding \cite{Liu}, and
its suitability to compare the performance of the proposed PD and
existing techniques. This metric is based on the channel
polarization tree and the number of paths on the tree with the same
number of zeros or ones, respectively. The channel polarization tree
is obtained by the recursive process of polarization channel
construction [1]. The branch of the tree obtained by
$Z(W^{(2i-1)}_n)$ in (\ref{index}) is labeled $1$ and the branch
obtained by $Z(W^{(2i)}_n)$ in (\ref{index}) is labeled $0$. Each
tree path refers to a polarized channel $W_i$ and to a row of the
generator matrix $\textbf{G}_n$.

The SD for path weight for the 1s is given by \cite{Niu}
\begin{equation}
d = \sum_{(k = 0)}^lP_1(l,k,Q)k = \sum_{(k = 0)}^l\frac{H_n^{(k)}}{n}k,
\end{equation}
where $P_1(l,k,Q)=\frac{H_n^{(k)}}{n}$ is the probability of path
weight $k$ with $Q=|\textbf{p}|$ bits shortening and $l$ refers to
the levels in the polarization tree.

The SD for path weight for the 0s is given by \cite{Niu}
\begin{equation}
\lambda = \sum_{(r = 0)}^lP_0(l,r,Q)r = \sum_{(r = 0)}^l\frac{C_n^{(r)}}{n}r,
\end{equation}
where $P_0(l,k,Q)=\frac{C_n^{(r)}}{n}$, where
$C_l^{(r)}=\binom{l}{r}$ is the probability of path weight $r$ with
$Q=|\textbf{p}|$ bits shortening and $l$ refers to the levels in the
polarization tree. The SD for path weight for the 0s used as the
main metric to evaluate the performance of the proposed and existing
shortening techniques.

The term $C(X)=\sum_{(r = 0)}^lC_l^{(r)}X^r$ describes the number of
paths with a given number of zeros, or alternatively
$C(X)=\sum_{(i=1:n)}X^{\rm Pb_i}$, $\rm Pb$ is the number of zeros
of each path. As an example, for a $\textbf{G}_{16}$, we have
$C(X)=X^0+4X^1+6X^2+4X^3+X^4$, one path with no zero, four paths
with 1 zero, six paths with 2 zeros, 4 paths with 3 zeros and one
path with 4 zeros, the $\lambda = \frac{1 \cdot 0 +4 \cdot 1+6 \cdot
2+4 \cdot 3+1 \cdot 4}{16}=2$.

Given a shortening procedure, the $C(X)$ is updated by removing the
paths cut by shortening, each path corresponds to a channel, which
in turn corresponds to a row (and column) in the generator matrix
$\textbf{G}$. For $\textbf{G}_{12}$ with $\textbf{p} = (14,15,16)$,
updating $C(X)=2X^1+5X^2+4X^3+1X^4$ and new $\lambda = \frac{2 \cdot
1+5 \cdot 2+4 \cdot 3+1 \cdot 4}{16}=1.75$, always less than the
previous value $\lambda$.

The set of shortened paths have different weights for each
shortening method, existing shortening ($\textbf{p}_{\rm ep}$) such
that $\textbf{p}_{\rm PD} \neq \textbf{p}_{\rm ep}$,
$|\textbf{p}_{\rm PD}|=|\textbf{p}_{\rm ep}|=\rm y>0$,  $\exists$
$\rm y$ we have
\begin{equation}
\sum_{i=1}^{\rm y}C_{\rm PD(i)}(X) \leq \sum_{i=1}^{\rm y}C_{\rm
CW(i)}(X), \label{ineq}
\end{equation}
with $C_{\rm PD(i)}(X)$ for PD technique and $C_{\rm CW(i)}(X)$ for
CW technique. 
We remark that there is a value of y from which the shortened
channels will be different. Expanding the above equation and
assuming that one chosen path of the PD set is different from that
of the CW set, we have
\begin{eqnarray}
\sum_{i=1}^{\rm y-1}C_{\rm PD(i)}(X) + \alpha X^{\rm y} <  
\sum_{i=1}^{\rm y-1}C_{\rm CW(i)}(X) + \beta X^{\rm y},
\end{eqnarray}
where $\alpha$ and $\beta$ are integer numbers. For small values of
y, where the channels shortened by either method will be the same,
we have the equality in (\ref{ineq}). We then exploit the fact that
$\alpha < \beta$, which yields
\begin{equation}
\alpha X^{\rm y} < \beta X^{\rm y},
\end{equation}
proving the inequality in (\ref{ineq}).

\section{Simulations}

In this section, we present simulations of rate-compatible polar
codes with shortening and a system equipped with the SC decoder, as
described for the Internet of Things (IoT) and the Enhanced Mobile
Broadband (eMBB) 5G scenarios \cite{Bioglio} and \cite{Ferreira},
which require the use of short to moderate block lengths. We note
that designs with Low-Density Parity-Check (LDPC) codes
\cite{bfpeg,dopeg,memd} and other decoding algorithms
\cite{mmimo,vfap,spa,jidf,mfsic,did,mbdf,bfidd} can also be studied.
We measure the BER and the frame error rate (FER) against the SNR,
defined as the ratio of the bit energy, $E_b$, and the power
spectral density, $N_0$, in dB. In the first example, we consider an
IoT scenario with $n=512$, $n'=480$ and $k=256$. In particular, we
have reduced the $\textbf{G}_{512}$ matrix to the $\textbf{G}_{480}$
matrix using the proposed PD shortening technique. For comparison
purposes, we have also included the curves associated with the best
performing existing methods CW and RQUP, as can be observed in Fig.
2. The results in Fig. 2 show that the proposed PD technique
outperforms the RQUP and the CW techniques by up to $0.25$ dB for
the same BER and FER performances, and approaches the performance of
the mother code (MC) with $n=512$ and $k=256$.

\begin{figure}[htb]
\begin{center}
\includegraphics[scale=0.48]{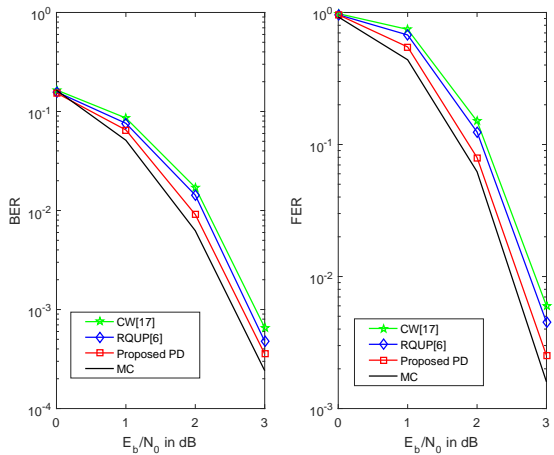}
\caption{BER and FER performaces of rate-compatible Polar Code
$n'$=480 $k$=256.}
\end{center}
\label{figura:fig03}
\end{figure}

In the second example, for eMBB scenario with $n=2048$, $n'=1920$
and $k=1600$. The results in Fig. 3 show that the proposed PD
technique outperforms the RQUP and the CW techniques by up to $0.20$
dB for the same BER and FER performances, and approaches the
performance of the mother code (MC) with $n=2048$ and $k=1600$.

\begin{figure}[htb]
\begin{center}
\includegraphics[scale=0.55]{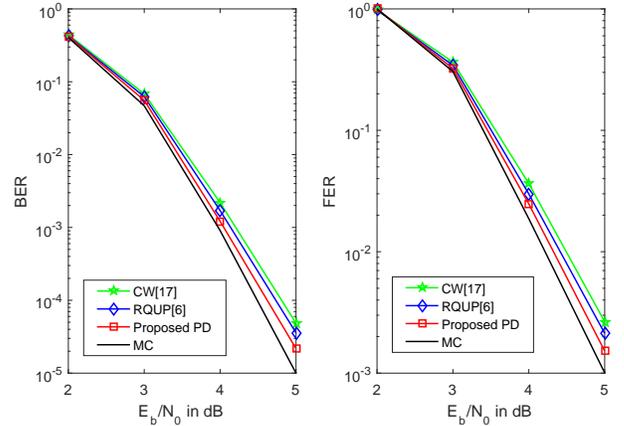}
\caption{BER and FER performaces of rate-compatible Polar Code
$n'$=1920 $k$=1600.}
\end{center}
\label{figura:fig04}
\end{figure}

In the Table II we compare for each simulation the SD values
obtained for each curve. Note that the SD of the proposed model PD
has a higher value than the CW and RQUP techniques.

\begin{table}[htb]
\caption{\label{tabela}Spectrum distance for Figs. 2 and 3}
\begin{center} \footnotesize
\begin{tabular}{|c||c|c|c|}\hline
\textit{$\rm SDC$}&Proposed PD &
RQUP\cite{Niu}&CW\cite{Wang}\\\hline \textit{$\rm Fig.
~2$}&4.53&4.46&4.43\\\hline \textit{$\rm Fig.
~3$}&5.46&5.44&5.43\\\hline
\end{tabular}
\end{center}
\end{table}

\section{Conclusion}

We have proposed the PD shortening method, which is based on the
channel polarization index, and can bring a performance improvement
in shortened polar codes as compared to existing shortening methods
in the literature. The use of the spectrum distance as a benchmark
for performance comparison has been shown as a valuable tool to
indicate the best shortening strategy, while requiring a low
computational complexity.

\end{document}